\documentclass[12pt]{article}
\usepackage{a4wide}
\usepackage{amssymb}
\usepackage{graphicx}
\begin{document}
{\renewcommand{\thefootnote}{\fnsymbol{footnote}}
\hfill  IGC--08/3--4\\
\medskip
\begin{center}
{\LARGE  Recollapsing quantum cosmologies\\ and the question of entropy}\\
\vspace{1.5em}
Martin Bojowald\footnote{e-mail address: {\tt bojowald@gravity.psu.edu}}
\\
\vspace{0.5em}
Institute for Gravitation and the Cosmos,
The Pennsylvania State
University,\\
104 Davey Lab, University Park, PA 16802, USA\\
\vspace{1.5em}
Reza Tavakol\footnote{e-mail address: {\tt r.tavakol@qmul.ac.uk}}
\\
\vspace{0.5em}
Astronomy Unit, School of Mathematical Sciences, 
Queen Mary, University of London, \\
Mile End Road, London, E1 4NS, UK
\vspace{0.5em}

\vspace{1.5em}
\end{center}
}

\setcounter{footnote}{0}

\newtheorem{theo}{Theorem}
\newtheorem{lemma}{Lemma}
\newtheorem{defi}{Definition}

\newcommand{\proofend}{\raisebox{1.3mm}{\fbox{\begin{minipage}[b][0cm][b]{0cm}
\end{minipage}}}}
\newenvironment{proof}{\noindent{\it Proof:} }{\mbox{}\hfill \proofend\\\mbox{}}
\newenvironment{ex}{\noindent{\it Example:} }{\medskip}
\newenvironment{rem}{\noindent{\it Remark:} }{\medskip}

\newcommand{\case}[2]{{\textstyle \frac{#1}{#2}}}
\newcommand{\lP}{\ell_{\mathrm P}}

\newcommand{\md}{{\mathrm{d}}}
\newcommand{\tr}{\mathop{\mathrm{tr}}}
\newcommand{\sgn}{\mathop{\mathrm{sgn}}}

\newcommand*{\R}{{\mathbb R}}
\newcommand*{\N}{{\mathbb N}}
\newcommand*{\Z}{{\mathbb Z}}
\newcommand*{\Q}{{\mathbb Q}}
\newcommand*{\C}{{\mathbb C}}

\begin{abstract}
  Recollapsing homogeneous and isotropic models present one of the key
  ingredients for cyclic scenarios. This is considered here within a
  quantum cosmological framework in presence of a free scalar field
  with, in turn, a negative cosmological constant and spatial
  curvature.  Effective equations shed light on the quantum dynamics
  around a recollapsing phase and the evolution of state parameters
  such as fluctuations and correlations through such a turn around. In
  the models considered here, the squeezing of an initial state is
  found to be strictly monotonic in time during the expansion, turn
  around and contraction phases.  The presence of such monotonicity is
  of potential importance in relation to a long standing intensive
  debate concerning the (a)symmetry between the expanding and
  contracting phases in a recollapsing universe. Furthermore, together
  with recent analogous results concerning a bounce one can extend
  this monotonicity throughout an entire cycle. This provides a strong
  motivation for employing the degree of squeezing as an alternative
  measure of (quantum) entropy. It may also serve as a new concept of
  emergent time described by a variable without classical analog. The
  evolution of the squeezing in emergent oscillating scenarios can in
  principle provide constraints on the viability of such models.
\end{abstract}
\section{Introduction}
Classical cosmology has shown enormous progress over the recent years.
Despite this a number of fundamental questions remain.  Central among
these is the fact that within the classical general relativistic
framework, the initial state of the universe is singular which would
result in the breakdown of laws of physics.  To obtain a satisfactory
scenario with a non-singular initial state one often looks to quantum
gravity and quantum cosmology. In fact, with a loop quantization one
can generically resolve the big bang singularity in cosmological and
other models \cite{LivRev,BSCG}. In the simplest cases, a bounce
results which keeps the volume non-zero and the universe away from the
classical singularity reached otherwise at the big bang.  The
possibility of a contracting phase (or several phases) before the hot
big bang has recently been invoked in a number of cosmological
scenarios, including several models proposed as alternatives to
standard inflation, such as for example pre-big bang \cite{PreBigBang}
and the ekpyrotic/cyclic scenarios
\cite{Ekpyrotic,CyclicEkpy,CyclicDesign}. The assumed nature of such
phases, however, has so far been mostly rather ad hoc, without a
satisfactory treatment of the classical singularity. The presence of
such phase(s) raises important questions, including their nature and
their relation to the present phase of the universe.  This in turn
relates to fundamental questions such as, among others, cosmological
entropy and the arrow of time.

Now given that the big bang was a high-energy, strong-curvature
regime, the understanding of the pre- and post-bounce phases would
require a full control of dynamical evolution of the quantum state
through such a bounce.  Moving through a bounce, a wave packet can
spread and deform significantly, implying that the universe before the
bounce could, for all we know, have been in a state very different
from what we see now. Thus, to understand the cosmological dynamics
through such bounces, all aspects of a quantum space-time are
essential, including its fluctuations and higher moments. 

In loop quantum cosmology, solvable models with controlled state
properties exist if the matter source is a free, massless scalar. This
has been analyzed numerically
\cite{QuantumBigBang,APS,APSII,APSCurved} and analytically
\cite{BouncePert,BounceCohStates}. More general models can be treated
by means of effective equations \cite{BouncePot,QuantumBounce}, as
they are also employed here for the recollapse.  Note that the concept
of effective equations is much more general than simply providing
correction terms to classical equations. With a complete set of
consistent effective equations one can, in fact, derive dynamical
properties such as expectation values, fluctuations, correlations or
higher moments for full quantum states. As we will see below, state
properties can be studied directly by using effective equations, which
provide an economical and representation-independent approximation
scheme of the evolution of states. (For another discussion of
effective equations especially in quantum cosmology, see
\cite{VaasLQCII}.)

If one combines the quantum bounce with a classical recollapse, cyclic
models ensue. Such oscillatory models, according to which the universe
undergoes many (and possibly an infinite number of) bounces, have been
employed in order to construct non-singular emergent models which can
set the initial conditions for a successful phase of inflation.  Since
such a universe can pass through many cycles, and hence many high
energy, strong-curvature regimes, this could result in even more
severe changes of its state compared to a single bounce.  We should
note that oscillatory models have a long history in cosmology at least
since the studies by Tolman in the 1930's \cite{TolmanEntropy,Tolman}
-- albeit within a classical setting. Interestingly, Tolman also
considered the question of cosmological entropy for these models,
claiming that the entropy during the expanding phase should be
slightly lower than during the subsequent collapsing phase. In these
studies entropy refers to that of the content of the universe
\cite{TolmanEntropy,Tolman} and ignores contributions from (quantum)
gravity.

There are, however, important problems with these models, including
the lack of treatment of singularities and the un-corroborated
assumption that the bounces themselves leave the entropy of the
universe unchanged.  The consideration of oscillatory models within a
quantum cosmological framework, on the other hand, not only allows
singularities to be avoided, but also introduces many more quantum
degrees of freedom, thus allowing the question of entropy to be
considered in a different light.

This is the setting we consider in this paper. We will analyze the
recollapse in detail, which is a semiclassical regime but,
crucially, still described in terms of a quantum state. We
especially focus on the evolution of state parameters through the
recollapse, which provides insights to the question of what their
generic change may be. In particular, we are interested in how
strongly fluctuations of a generic state respect time-reversal
symmetry for time reflections around the recollapse point. If
fluctuations are symmetric in this sense, there is not much change
between the pre- and post-recollapse phases. A violation of the
symmetry, on the other hand, would provide a measure for the generic
change of the quantum state in the recollapse phase.  The analysis is
thus complementary to what has already been studied for the bounce
\cite{BeforeBB,Harmonic}: Can the quantum state after the recollapse
be very different from what it was before? Especially in the presence
of many cycles, this question is important for understanding the
viability of oscillatory cosmological models over epochs long
compared to the life time of individual cycles.

For technical reasons, we shall take a free massless scalar as the
matter source in all models considered in detail here. However, we
shall also demonstrate the robustness of our claims under the
inclusion of potentials.  The free scalar has the advantage that it
can be used as a global internal time parameter and thus gives rise to
true Hamiltonian, rather than constrained, evolution. Any non-constant
potential or even a mass term would spoil this feature.  (Here we
refer to the classical situation. We will later encounter and
entertain the possibility of genuine quantum variables as a measure
for time even in situations where no obvious classical clock may
exist.) Moreover, in the absence of a cosmological constant and for
flat, isotropic space, this matter content provides an exactly
solvable model even after quantization (loop or otherwise)
\cite{BouncePert}. Thus, there are no dynamical quantum corrections
whatsoever in this case; the system is harmonic and presents the
simplest and most controlled model of quantum cosmology. (There may,
however, be quantum geometry corrections of kinematical type which
give rise to a bounce in loop quantum cosmology. But they turn out not
to spoil the dynamical solvability \cite{BouncePert}.)  However, this
exact model does not allow a recollapse, and we therefore have to add
extra ingredients and with them non-trivial quantum
corrections. Nevertheless, the resulting systems will be manageable
and provide key contributions for highly controlled cyclic
models. While there is no scalar potential in the main part of the
paper, we verify that in fact our results remain robust in
presence of general non-zero potentials. Moreover, our analysis
provides a starting point to analyze equations in the presence of a
potential perturbatively. For the bounce, such equations are developed
in \cite{BouncePot,QuantumBounce}, which in some cases even allow
conclusions valid to all orders in the potential and in quantum
moments \cite{BounceSqueezed}. Since our main question is about
limitations to the symmetry of fluctuations around cosmological
turning points, a highly controlled model is reliable as any
limitation there would only grow if the model becomes more
complicated. (See also \cite{BeforeBB,Harmonic} in this context.)  In
Sec.~\ref{s:SpreadEvol} we will comment in more detail on possible
effects of a potential.

\section{Recollapsing models}
We shall confine ourselves to isotropic and homogeneous settings.
There are two different ways to achieve a recollapsing cosmological
model: by including a negative cosmological constant or by allowing
positive spatial curvature. We shall first describe the general scheme
of our analysis and then specialize to these two cases.
\subsection{Prescription}
In the presence of a cosmological constant and a free massless scalar
field the Friedmann equation takes the form
\begin{equation}
 \left(\frac{\dot{a}}{a}\right)^2+\frac{k}{a^2}= \frac{4\pi
 G}{3}\frac{p_{\phi}^2}{a^6}+\Lambda , 
\end{equation}
where $p_{\phi}$ is the momentum corresponding to the homogeneous scalar
field $\phi$ and can be written as
\begin{eqnarray}
 p_{\phi}&=&\pm a^2 \sqrt{\frac{3}{4\pi G}}\sqrt{\dot{a}^2+k-\Lambda a^2}\nonumber\\
& =&\pm
 2\sqrt{\frac{8\pi G}{3}} (1-x)V\\
&&\times\sqrt{P^2+kf_0^2\left(\frac{8\pi G(1-x)f_0V}{3}\right)^{\frac{2x}{1-x}}-
\Lambda f_0^2 \left(\frac{8\pi  G(1-x)f_0V}{3}\right)^{\frac{1+2x}{1-x}}} ,
\nonumber
\end{eqnarray}
in terms of canonical gravitational variables
\begin{equation}
 V=\frac{3a^{2-2x}}{8\pi  G(1-x)f_0} \mbox{ and }P=-f_0a^{2x}\dot{a} 
\quad\mbox{with}\quad \{V,P\}=1\,.
\end{equation}
The introduction of this pair of canonical variables and the
parameters $f_0$ and $x$ is motivated by loop quantum cosmology and
deserves further explanation:
While $f_0$ will not be of much consequence in what follows, we keep
it for general reference. It has dimensions such that $P$ becomes
dimensionless; for $x=-1/2$, for instance, it has the dimension of
length and for $x=0$ it is itself dimensionless. Its significance lies
in the fact that it determines a fundamental scale for the loop
quantization which becomes relevant at the bounce. Moreover, the
fundamental length can depend on the evolution of the universe, and
thus $a$, if the underlying discreteness of quantum gravity is being
refined during evolution. This possibility is taken into account by
the parameter $x$, which makes the momentum $P$ depend differently on
$a$ for different choices of $x$. How precisely these parameters arise
has been discussed, e.g., in \cite{InhomLattice,SchwarzN}. The
dynamical behaviour of loop quantum cosmology is sensitive to their
values, but in this paper we will mainly analyze recollapses where
effects of the loop quantization are not expected to play large roles.
We will nevertheless see that it is of interest to keep all
possibilities, especially of $x$. For all choices of $f_0$ and $x$,
the variables used here are canonically related to each other.
Nevertheless, some quantitative aspects can change, and also equations
of motion may be easier to solve for some $x$ than others.

Physically, different values of $x$ correspond to different ways in
which an inhomogeneous discrete quantum state can be refined during
its evolution on microscopic levels \cite{InhomLattice}.  For $x=0$
the variable $P$ corresponds to an underlying state which has a
constant number of lattice sites as the universe expands, while for
$x=-1/2$ the state has a constant geometrical size at each lattice
site and thus requires new sites to be generated during expansion. A
precise value of $x$ could in principle be determined if one could
derive a reduced Hamiltonian of an isotropic model from a full,
inhomogeneous Hamiltonian (such as those introduced in
  \cite{QSDI}).  Since this is not yet available, we have to keep the
the value of $x$ free and look instead for possible
phenomenological constraints.

Classical solutions as functions of $\phi$ are readily determined from
the Hamiltonian $H\propto p_{\phi}$ and its canonical equations of
motion in terms of $\phi$, which will be presented below. Such
equations of motion determine the relational dependence of, e.g.,
$V(\phi)$ through the Hamiltonian equation of motion $\md V/\md\phi=
\{V,H\}$. Our main interest, however, is in possible effects which may
result from the behaviour of quantum states. In particular, a quantum
system has not only expectation values as free variables, which could
be associated with the classical variables $(V,P)$, but also
fluctuations, correlations and higher moments. Dynamically, all these
variables couple in a general quantum system. These coupled equations
of motion can be derived from the usual commutator relations uch as
$\md\langle\hat{V}\rangle/\md\phi=
-i\hbar^{-1}\langle[\hat{V},\hat{H}]\rangle$ or, more compactly, from
a quantum Hamiltonian $H_Q:=\langle\hat{H}\rangle$.  (For details we
refer to \cite{EffAc,Karpacz} or, in the context of cosmological
models, \cite{BounceCohStates,Harmonic}.) Here, the expectation value
is computed in a state with a general set of moments. As is well
known, for a general classical Hamiltonian $H(V,P)$ we have $\langle
H(\hat{V},\hat{P})\rangle\not=
H(\langle\hat{V}\rangle,\langle\hat{P}\rangle)$ where the difference
amounts to quantum corrections to the classical dynamics.  These
corrections depend, e.g., on quantum fluctuations or, more generally,
on moments
\begin{equation}
 G^{a,b}=\langle(\hat{V}-\langle\hat{V}\rangle)^a
 (\hat{P}-\langle\hat{P}\rangle)^b\rangle_{\rm Weyl} ,
\end{equation}
of the state used for the expectation values. (In the definition of
moments, we assume the basic operators to be totally symmetric or Weyl
ordered as indicated by the subscript.) Upon writing
$H_Q=\langle\hat{H}\rangle$ in terms of expectation values and the
moments, we obtain the complete quantum Hamiltonian.  This in turn
generates the Hamiltonian equations of motion for\footnote{From
    now on, we will mostly be referring to the quantum theory unless
    stated otherwise, and thus drop brackets on expectation values.}
$V:=\langle\hat{V}\rangle$, $P:=\langle\hat{P}\rangle$ as well as all
the moments $G^{a,b}$. (As before, the equations of motion are
  given by $\md f/\md\phi=\{f,H_Q\}$ where $\{V,P\}=1$ and for $G^{a,b}$
the Poisson brackets follow from expectation values of commutators
divided by $i\hbar$.)

This is the basis for the derivation of effective equations which may
provide good approximations in regimes where the infinite set of all
moments can be truncated to finitely many variables. In the following
we shall only consider the second order moments which, for better
clarity, we denote as
\begin{eqnarray}
 G^{PP} &=& G^{0,2}=\langle\hat{P}^2\rangle-P^2\\
 G^{VP} &=& G^{1,1}=\frac{1}{2}\langle\hat{V}\hat{P}+\hat{P}\hat{V}\rangle 
 -VP\\
 G^{VV} &=& G^{2,0}=\langle\hat{V}^2\rangle-V^2\,.
\end{eqnarray}
Their Poisson brackets can be then derived as in
\begin{eqnarray}
 \{G^{VV},G^{PP}\} &=& \{\langle\hat{V}^2\rangle-V^2,
 \langle\hat{P}^2\rangle-P^2\}\nonumber\\
&=& \frac{1}{i\hbar} \langle[\hat{V}^2,\hat{P}^2]\rangle- 
\frac{2P}{i\hbar}\langle[\hat{V}^2,\hat{P}]\rangle- 
\frac{2V}{i\hbar}\langle[\hat{V},\hat{P}^2]\rangle+ 
\frac{4VP}{i\hbar}\langle[\hat{V},\hat{P}]\rangle\nonumber\\
&=& 2\langle\hat{V}\hat{P}+\hat{P}\hat{V}\rangle- 4VP= 4G^{VP}\,. \label{PBG1}
\end{eqnarray}
Similarly,
\begin{equation} \label{PBG2}
 \{G^{VV},G^{VP}\}= 2G^{VV} \quad\mbox{and}\quad \{G^{VP},G^{PP}\}=
 2G^{PP}\,.
\end{equation}
Such Poisson brackets, when used in $\md G^{a,b}/\md\phi =
\{G^{a,b},H_Q\}$, determine the evolution of the quantum variables of a
state. This demonstrates how effective equations are able to go well
beyond simple corrections to classical equations, which will be made
ample use of in this article.

\subsection{Negative cosmological constant}

For $\Lambda<0$, $k=0$, our system has the classical Hamiltonian
\begin{equation}
  H=(1-x)V\sqrt{P^2+|\Lambda|f_0^2 (8\pi\gamma G(1-x)f_0V/3)^{(1+2x)/(1-x)}},
\end{equation}
for $\phi$-evolution, i.e.\ $p_{\phi}=2\gamma \sqrt{8\pi G/3}H$
  (a specific sign has been chosen here for the square root; the other
  choice simply amounts to replacing $\phi$ with $-\phi$). The factor
in $p_{\phi}$ can be eliminated by redefining $\phi$.  Evolution is
analyzed best for $x=-1/2$, in which case
\begin{equation}
 H= \frac{3}{2}V\sqrt{P^2+|\Lambda|f_0^2},
\end{equation}
is linear in $V$. The corresponding quantum Hamiltonian, including
moments of second order, is
\begin{equation} \label{HQ}
 H_Q = \frac{3}{2}V\sqrt{P^2+|\Lambda|f_0^2}+ \frac{3}{4} |\Lambda|f_0^2
\frac{V}{(P^2+|\Lambda|f_0^2)^{3/2}} G^{PP}+ \frac{3}{2}
\frac{P}{\sqrt{P^2+|\Lambda|f_0^2}} G^{VP}, 
\end{equation}
which includes the quantum moments $G^{PP}$, $G^{VP}$ in correction
terms. Higher moments are ignored here, and $G^{VV}$ does not occur
thanks to the linearity of $H$ in $V$. (For $\Lambda=0$ we have the
solvable free system, in which no coupling terms between expectation
values and moments arise \cite{BouncePert}.)  The quantum Hamiltonian
determines the Hamiltonian equations of motion
\begin{eqnarray}
 \frac{\md V}{\md\phi} &=& \frac{3}{2}\frac{VP}{\sqrt{P^2+|\Lambda|f_0^2}}
-\frac{9}{4}|\Lambda|f_0^2 \frac{VP}{(P^2+|\Lambda|f_0^2)^{5/2}} G^{PP}
+\frac{3}{2}|\Lambda|f_0^2 \frac{G^{VP}}{(P^2+|\Lambda|f_0^2)^{3/2}} 
\label{Veom}\\ 
\frac{\md P}{\md\phi} &=& -\frac{3}{2}\sqrt{P^2+|\Lambda|f_0^2}- \frac{3}{4}|\Lambda|f_0^2
\frac{G^{PP}}{(P^2+|\Lambda|f_0^2)^{3/2}}\,. \label{Peom}
\end{eqnarray}

Quantum fluctuations appear here in coupling terms and are themselves
dynamical, subject to equations of motion
\begin{eqnarray}
 \frac{\md G^{PP}}{\md\phi} &=& -3\frac{P}{\sqrt{P^2+|\Lambda|f_0^2}} G^{PP} 
\label{GPPeom}\\
 \frac{\md G^{VP}}{\md\phi} &=& \frac{3}{2}|\Lambda|f_0^2
\frac{V}{(P^2+|\Lambda|f_0^2)^{3/2}}
 G^{PP} \label{GVPeom}\\
 \frac{\md G^{VV}}{\md\phi} &=& 3 |\Lambda|f_0^2
\frac{V}{(P^2+|\Lambda|f_0^2)^{3/2}} G^{VP}
 +3\frac{P}{\sqrt{P^2+|\Lambda|f_0^2}} G^{VV}\,. \label{GVVeom}
\end{eqnarray}
These equations satisfy
\[
 \frac{\md}{\md\phi}\left(G^{VV}G^{PP}-(G^{VP})^2\right)=0
\]
such that a state initially saturating the (generalized) uncertainty
relation
\begin{equation}\label{uncert}
 G^{VV}G^{PP}-(G^{VP})^2\geq \frac{\hbar^2}{4}
\end{equation}
will keep saturating it. Such a state would be considered a dynamical
coherent state whose properties can be analyzed by our equations. In
what follows, however, we will not restrict states to be on the
saturation surface although they certainly must satisfy the
uncertainty relation.

If we first ignore all moments and their quantum back-reaction, we
find the classical solutions
\begin{eqnarray}
 P_{\rm classical}(\phi) &=& P_0 \cosh(3(\phi-\phi_0)/2)+
 \sqrt{P_0^2-|\Lambda|f_0^2} \sinh(3(\phi-\phi_0)/2)\label{Pclass}\\ V_{\rm
 classical}(\phi) &=& V_0\frac{\sqrt{P_0^2+|\Lambda|f_0^2}}{-P_0
 \sinh(3(\phi-\phi_0)/2)+ \sqrt{P_0^2+|\Lambda|f_0^2}
 \cosh(3(\phi-\phi_0)/2)}\,.\label{Vclass}
\end{eqnarray}
The volume has a turning point, and we can simplify expressions
without loss of generality by choosing our initial values there, i.e.\
$P_0=P(\phi_0)=0$ and shift $\phi$ such that $\phi_0=0$. Then, we have
simply
\begin{eqnarray}
 P_{\rm classical}(\phi) &=& -\sqrt{|\Lambda|}f_0 \sinh(3\phi/2)
 \label{Pclasssimple}\\ 
V_{\rm  classical}(\phi) &=& \frac{V_0}{\cosh(3\phi/2)}\,.
\label{Vclasssimple}
\end{eqnarray}
These solutions describe the recollapse of a universe with a past and
  a future singularity. Analytical solutions of equations amended by
  quantum geometry effects, where the singularities are replaced by
  bounces and thus provide cyclic solutions, have been derived e.g.\
  in \cite{BounceSols}. However, quantum back-reaction effects, which
  complicate the analysis, were not included in the equations used
  there.

In a next step, we can solve the equations of motion
(\ref{GPPeom})--(\ref{GVVeom}) approximately by assuming the classical
solutions for $P$ and $V$. Thus, we are still ignoring quantum
back-reaction effects at this stage, which if present would imply that
the moments back-react by the coupling terms in (\ref{Veom}) and
(\ref{Peom}) and change the classical solutions. For small
fluctuations, this will be a good approximation, and solutions
  obtained for the moments will allow us to check self-consistently
for how long in $\phi$ it will remain valid.

It is then easy to solve for $G^{PP}$, to give
\begin{equation}
 G^{PP}(\phi) = G_0^{PP}\cosh^2(3\phi/2) ,
\end{equation}
which shows that $G^{PP}$ is inversely proportional to the volume squared,
and which in turn allows to solve for
\begin{equation} \label{GVPsol}
 G^{VP}(\phi) = G_0^{VP}+\frac{V_0G_0^{PP}}{\sqrt{|\Lambda|}f_0}
 \frac{\sinh(3\phi/2)}{\cosh(3\phi/2)}\,.
\end{equation}
With this, one can finally solve for
\begin{equation} \label{GVVsol}
G^{VV}= \frac{G_0^{VV}+2\frac{V_0G_0^{VP}}{\sqrt{|\Lambda|}f_0}
 \tanh(3\phi/2)+
  \frac{V_0^2G_0^{PP}}{|\Lambda|f_0^2}
  \tanh^2(3\phi/2)}{\cosh^2(3\phi/2)}\,.
\end{equation}

With quantum back-reaction to second order in moments, i.e.\ solving
the full equations (\ref{Veom})--(\ref{GVVeom}) without starting
  with the classical solutions, the equations are more highly
coupled. One can derive some solutions by dividing (\ref{Peom}) by
(\ref{GPPeom}), thus providing a differential equation for
$P(G^{PP})$:
\begin{equation}
\frac{\md P}{\md G^{PP}} = \frac{P^2+|\Lambda|f_0^2}{2PG^{PP}}+
\frac{1}{4}\frac{|\Lambda|f_0^2}{P(P^2+|\Lambda|f_0^2)}\,.
\end{equation}
This can be written in a simpler form thus
\begin{equation}
 \frac{\md (P^2+|\Lambda|f_0^2)}{\md \log G^{PP}} = P^2+|\Lambda|f_0^2+
 \frac{1}{2}\frac{\Lambda}{P^2+|\Lambda|f_0^2} G^{PP},
\end{equation}
whose solution yields
\begin{equation}
 P = \sqrt{-|\Lambda|f_0^2+\sqrt{c (G^{PP})^2-
     |\Lambda| G^{PP}}}
\end{equation}
such that
\begin{equation}
 \sqrt{P^2+|\Lambda|f_0^2} = \sqrt[4]{c(G^{PP})^2-
   |\Lambda|f_0^2 G^{PP}},
\end{equation}
with a constant of integration $c$.

\subsection{Positive spatial curvature}

With $\Lambda=0$ but $k=1$, the system is simplest to solve for $x=0$,
which makes it again linear in $V$. The quantum Hamiltonian is then
the same as before, (\ref{HQ}), with $\Lambda$ replaced by $-1$ (and a
missing factor of $3/2$ arising from $1-x$ in the Hamiltonian, which
simply rescales $\phi$).  We can thus immediately take over the
solutions already found. For other values of $x$, the equations are
more highly coupled and do not allow simple solutions. Nevertheless,
we can use the solutions already provided to find information also
about these systems by simply replacing $(V,P)$ in the $x=0$-solutions
by
\begin{equation} \label{tilde}
 \tilde{V}:=\frac{1}{Gf_0}((1-x)Gf_0V)^{1/(1-x)}\quad,\quad
\tilde{P}:=\frac{P}{((1-x)Gf_0V)^{x/(1-x)}}=P/(Gf_0\tilde{V})^{x}\,.
\end{equation}
(We have chosen the factors of $G$ and $f_0$ such that $\tilde{P}$ has
the same dimensions as $f_0$, which will be useful later.)  This has
to be done also in the moments, i.e.\ we will obtain their solutions
not for $G^{VV}$, say, but for $G^{\tilde{V}\tilde{V}}$. These are not
directly the fluctuations of our basic variables for $x\not=0$ but
they still give important information about the spreading and other
properties of states. For instance, we will determine the correlation
$G^{\tilde{V}\tilde{P}}$ instead of $G^{VP}$. Both parameters contain
equally interesting information about squeezing and the symmetry of
fluctuations around the recollapse. In particular, if
$G^{\tilde{V}\tilde{V}}$ is not symmetric around the recollapse, then
nor will be $G^{VV}$.

\section{Implications}

Several conclusions can be drawn from the solutions found to the given
order.

\subsection{Volume ratio between recollapse and high curvature
  regimes}

Our solutions correspond to state parameters in a Wheeler--DeWitt
quantization because we use elementary variables $(V,P)$ which are
assumed to be quantized to well-defined operators. Those operators,
together with the Hamiltonian, then determine the dynamics. The latter
have not been written explicitly here, but they are the central
ingredient to Hamiltonian equations of motion via the Poisson brackets
of quantum variables such as (\ref{PBG1}) and (\ref{PBG2}).

The Wheeler--DeWitt quantization does not easily solve the singularity
problem. For models without quantum back-reaction effects, i.e.\
spatially flat models sourced by a free massless scalar,
$\langle\hat{V}\rangle$ simply follows the classical trajectory into
the singularity. On the other hand, in general models such as those
considered here, there are quantum back-reaction effects which one may
expect to become stronger as the solution for $V$ approaches zero ---
the classical singularity. This could stop $V$ altogether, or delay
its approach to zero sufficiently strongly such that zero would not be
reached in a finite amount of proper time (but possibly still finite
in $\phi$). However, this is difficult to analyze if all moments are
required, and unlikely to result in a generic resolution of
singularities.

A loop quantization does provide a natural solution of the singularity
problem in isotropic models, but it requires one to use a different
set of basic variables. (At a basic level, singularities in
homogeneous and spherically symmetric models have been shown to be
absent by allowing general wave functions to be extended through
classical singularities \cite{Sing,HomCosmo,Spin,SphSymmSing}. More
specific examples for bouncing wave packets are derived in
\cite{QuantumBigBang,BouncePert}. For a discussion and comparison of
results concerning singularities see \cite{BSCG}.) While $V$ would
still be represented as an operator in the quantization, the curvature
(or connection) component $P$ is not. Instead, loop quantum gravity is
based on a quantum representation in which only holonomies of the
Ashtekar connection are represented, in this way providing the
kinematical structures for a well-defined, background independent
quantization of full gravity \cite{Rov,ALRev,ThomasRev}. In the
cosmological models studied here, this means that it is not $P$ which
is part of the elementary algebra but $\exp(i\mu P)$, for arbitrary
real $\mu$. (Note it is $P$ which enters here, rather than $\tilde{P}$
of (\ref{tilde}) because $x$ represents the freedom in the refinement
of a discrete underlying state and thus determines the form of
holonomies in a reduced isotropic setting \cite{InhomLattice}. This
is, in fact, the main reason why we allow for different values of
$x$.) Using the exponential instead of an expression linear in
  $P$ changes the basic algebra as well as the Hamiltonian in
particular at large $P$. In a flat, isotropic model with a free scalar
field, the classical singularity is then resolved and replaced by a
bounce.

To study the oscillating models we need to consider a combination of
bounces and recollapses which is more complicated because of the
structure of required quantum evolution equations. Nevertheless, one
can study cyclic solutions by patching together bounce and recollapse
phases. For small curvatures, we can use the equations and the
corresponding solutions provided in this paper to an excellent
approximation, even for a model of loop quantum cosmology. However, we
can use this only when $P$ is not too large and have to cut off our
solutions at the latest when $|P|\sim 1$. (At this point, the
  precise value of $f_0$ would set the corresponding scale for
  $\dot{a}$.) This leaves only a finite range of sizes for the
universe between this high curvature regime and the recollapse. The
high curvature regimes can also be described by effective
equations, which are in fact precise without quantum back-reaction,
but require a different set of basic variables \cite{BouncePert}.

For a negative cosmological constant, we have the ratio
$V_0/V_{|P|=1}= \sqrt{1+1/|\Lambda|f_0^2}$. Thus, for a small
cosmological constant compared to $f_0^{-2}$, the ratio is huge. Since
$f_0$ arises from quantum gravity and has the dimension of length in
this case which is based on $x=-1/2$, $f_0$ should take a value near
the Planck length. Thus, $|\Lambda|$ must only be small compared to a
Planckian value which can safely be assumed to be the case. For the
closed model with $x=0$, on the other hand, we have
$V_0/V_{|P|=1}=\sqrt{1+1/f_0^2}$ with a dimensionless $f_0$. In this
case, there are no strong reasons to expect quantum gravity to provide
a value of $f_0$ small compared to one (without reference to a second
scale larger than the Planck length, which should not appear in the
basic variables $V$ and $P$ where $f_0$ enters).  This is certainly
not enough for a macroscopic universe which has to grow large out of
the high curvature regime. For this reason we have to use other values
for $x$ in this case: then, $|P|=1$ is reached at much smaller values
for $\tilde{P}$ as provided by our solutions. Although the qualitative
behaviour is unchanged compared to other $x$, changes in $x$
  have an important quantitative implication (which was first
emphasized in \cite{APSCurved}). For example for $x=-1/2$, the
high curvature regime starts at
\[
\cosh(\phi)\sim \frac{1}{6}\,\sqrt[3]{108\,C+12\,\sqrt
{-12+81\,{C}^{2}}}+ \frac{2}{\sqrt[3]{108\,C+12\,\sqrt
{-12+81\,{C}^{2}}}}
\]
where $C=(Gf_0V_0)^{2/3}/f_0^2$.  For large $V_0$, this is
approximately $\cosh(\phi)\approx V_0^{2/9}$ (or, for general
$x\not=0$, $\cosh(\phi)\approx V_0^{-x/(1-x)^2}$). Thus, the ratio
$V_0/V_{|P|=1}\approx V_0^{-x/(1-x)^2}$ is no longer constant and grows
with $V_0$ for negative $x$. For $x=-1/2$, the ratio is given by
$V_0^{2/9}$ which is large enough for large $V_0$, leaving ample room
for a growing universe.
\subsection{Quantum back-reaction effects}
From our solutions we can determine whether quantum back-reaction
effects are strong around the recollapse. As one can easily see, there
are no possible divergences in the equations of motion (\ref{Veom})
and (\ref{Peom}) which would enhance the coupling terms. Quantum
back-reaction effects can only be strong if the quantum variables are
large, which can be avoided at least for some time by choosing a
semiclassical initial state. Thus, the equations to the order provided
here are reliable to a high degree and can be used to determine the
state properties around the recollapse. In particular, our equations
of motion and solutions for quantum variables themselves can be used
to see how long the approximation remains valid.
\subsection{Evolution of the spread}
\label{s:SpreadEvol}
Of particular interest is whether fluctuations depend strongly on
$\phi$ or remain nearly constant during the evolution. If they change
rapidly, the behaviour of neighbouring cycles would be noticeably
different from each other because the state would have changed
significantly. In scenarios with a large or an infinite number 
of cycles, large differences should even be generic between widely 
separated cycles.

As we have seen, $G^{PP}$ is always proportional to the inverse volume
squared when quantum back-reaction effects can be ignored. Thus, curvature
fluctuations must be symmetric around the recollapse and do not change
significantly: At any volume after the recollapse we have the same
$G^{PP}$ as at the same volume before.  For the other quantum
variables, however, the situation is different. Ignoring products of
quantum variables, we can rewrite (\ref{GVVeom}) approximately as
\begin{equation} \label{GVPVLambda}
 \frac{\md}{\md\phi}\left(\frac{G^{VV}}{V}\right) =
 3|\Lambda|f_0^2 \frac{G^{VP}}{(P^2+|\Lambda|f_0^2)^{3/2}}, 
\end{equation}
for a negative cosmological constant with $x=-1/2$ or
\begin{equation} \label{GVPVClosed}
 \frac{\md}{\md\phi}\left(\frac{G^{VV}}{V}\right) =
 \frac{2G^{VP}}{(P^2+f_0^2)^{3/2}}, 
\end{equation}
for a closed model with $x=0$ and $\Lambda=0$. This shows that
$G^{VV}$ would be a function only of $V$, and thus symmetric around
the recollapse, if $G^{VP}=0$, i.e.\ the state is unsqueezed. One may
assume this as an initial condition, but $G^{VP}$ itself is dynamical
and subject to the evolution equation (\ref{GVPeom}). Its time
derivative cannot be zero since, thanks to the uncertainty relation,
$G^{PP}$ is non-zero unless volume fluctuations diverge. Even an
initially unsqueezed state will become squeezed after some time, and
thus also affect the volume fluctuations.

Even if $G^{VV}/V$ is not constant, $G^{VV}$ may be symmetric
  around the recollapse but behave differently with respect to $V$. In
  fact, (\ref{GVVsol}) shows that $G^{VV}$ is symmetric around the
  recollapse if $G_0^{VP}$, i.e.\ the correlation at the recollapse,
  vanishes even though $G^{VP}$ would become non-zero away from the
  recollapse. But since this happens only under the special condition
  of $G_0^{VP}=0$, it could generically be satisfied only in one cycle
  of an oscillatory universe.

From (\ref{GVPsol}), we can estimate the change in squeezing per
recollapse by
\begin{equation}\label{SqueezingChange}
 \lim_{\phi\to\infty} G^{VP}(\phi)- \lim_{\phi\to-\infty}G^{VP}(\phi)
 = \frac{2 V_0G_0^{PP}}{\sqrt{|\Lambda|}f_0}
\end{equation}
as an upper bound. The change may be small for small fluctuations
$G_0^{PP}$, but is enlarged by a factor of $V_0$ (as well as
$1/\sqrt{|\Lambda|}f_0$ in the presence of $\Lambda<0$, which is large
given that $|\Lambda|f_0^2$ is small; if the recollapse is
  triggered by positive spatial curvature, we have the same formula
  with $\Lambda$ set to $-1$). In a large universe, this change can
be quite significant.  Note that in (\ref{SqueezingChange}) we have
used $\phi\to\pm\infty$, and thus a range which includes the high
curvature regimes where the equations have to be amended by effects of
the loop quantization and the specific solution would change. We
can take this into account by reducing the range of $\phi$; however,
this does not change the result but only affects the numerical factor
in the change of squeezing. There is thus a significant change during
the classical recollapse, irrespective of how the high curvature
regime is dealt with.  For instance, we have
\begin{equation}
 G^{VP}|_{\sinh(3\phi/2)=1}= G_0^{VP}+
  \frac{V_0G_0^{PP}}{\sqrt{2|\Lambda|}f_0}\,,
\end{equation}
whose numerical coefficient is different, but which still carries the
large factor of $V_0$. In fact, the $\tanh$-behaviour of $G^{VP}$
  demonstrates that the greatest change in correlations occurs near
  the recollapse.

 To quantify the production of squeezing during
  recollapse phases, it may be helpful to transform the solution for
  $G^{VP}(\phi)$ to proper time rather than using the relational
  formulation with respect to $\phi$. The relation between proper time
  $\tau$ and $\phi$ can in general be complicated, but can easily be
  obtained for $x=-1/2$ by integrating
\[
 \frac{\md\phi}{\md\tau} = \frac{p_{\phi}}{V}=
 \frac{p_{\phi}}{V_0}\cosh(3\phi/2)
\]
to obtain
\[
 \phi(\tau) = \frac{2}{3}{\rm arsinh}(\tan (3p_{\phi}(\tau-\tau_1)/2V_0))\,.
\]
Without loss of generality, we chose $\phi$ to vanish at $\tau_1$,
which may be different from the recollapse time $\tau_0$.  The whole
range $-\infty<\phi<\infty$ corresponds to a finite proper time
interval $-V_0\pi/3p_{\phi}<\tau-\tau_1<V_0\pi/3p_{\phi}$. This
highlights the fact that we are not including effects of the loop
quantization, such that the endpoints of the $\phi$-range, where the
volume vanishes, correspond to future and past singularities a finite
proper time away.

Inserting this in the solution (\ref{GVPsol}) for $G^{VP}$, we obtain
\begin{equation}
 G^{VP}(\tau)-G^{VP}(\tau_0) = \frac{V_0G_0^{VP}}{\sqrt{|\Lambda|}f_0}
 \sin (3p_{\phi}(\tau-\tau_1)/V_0)\,,
\end{equation}
which shows the growth of squeezing in proper time during each
recollapse (which is in fact monotonic in the given range of $\tau$).

Starting with an initially unsqueezed state it may seem that for many
cycles the state remains almost unchanged from cycle to cycle. Its
volume fluctuations may always seem to attain nearly the same size at
the same volume. However, this is so only because of the special
initial state chosen, from which squeezing builds up slowly. For small
$G^{VP}$, (\ref{GVPVLambda}) and (\ref{GVPVClosed}), respectively,
show that $G^{VP}/V$ is nearly constant in both cases considered.  The
change in volume fluctuations before compared to after the
  recollapse seems insignificant from cycle to cycle but becomes
noticeable over many cycles. Moreover, if the initial state had
already had some squeezing, volume fluctuations relative to volume
would change much more rapidly. In this way, the choice of initial
state can strongly influence the long-term behaviour.

In a cyclic model, it is especially important to ask what significance
one should attribute to the choice of initial state. Is it to be posed
in ``our'' cycle, and if not, how many cycles ago? If we could have
observational input on properties of the state, we could certainly
pose an initial condition in our cycle and see how the state evolves
to or from there. However, state properties are hardly under control,
and this possibility remains elusive. We thus have to pose initial
conditions many cycles ago based on some general principle of
emergence, but we never know how many. Thus, even though we know that
an initially unsqueezed state builds up squeezing only slowly, this
does not say much about the present state if we do not know how many
cycles ago the state was unsqueezed.

An interesting question is whether in a cyclic model one generically
expects to have a finite or an infinite number of past cycles. The
problem with the finite case is that it does not resolve the origin
question. In the emergent scenarios
\cite{Emergent,Emergent2,EmergentLoop,EmergentNat,EinsteinStaticHol},
as well as some other such models, the universe is assumed to have
undergone an infinite number of past cycles so as to remove the
question of the origin.  In that case any given cycle would have an
infinite number of precursors and generically we therefore have to
expect the current state to be squeezed. (We will argue in the next
subsection that bounces do not affect the qualitative behaviour of the
squeezing, especially its monotonicity).  The question then is how the
squeezing in a generic cycle is determined.  If each cycle produces
the same amount of squeezing, a generic cycle would have infinitely
squeezed states, which could not be semiclassical. However, as
(\ref{SqueezingChange}) shows, the amount of new squeezing per cycle
depends on the recollapse volume $V_0$ of that cycle. For growing
cycles, as in the emergent scenario, the change in squeezing is
initially small and approaches zero for cycles in the infinitely
distant past. Depending on the precise scenario, the sum of all
squeezing contributions may converge, such that a finite value results
for a generic cycle. Whether this is the case and what this precise
value could be depends on which concrete model one is using, and we
will not follow this route here. It is, however, interesting that this
in principle allows one to restrict the possibilities for emergent
scenarios by the amount of squeezing they would predict.

Another interesting and related question is that in the emergent
models the eventual non-uniformity of cycles is produced by a
non-constant potential. (In initial regions where the potential is
flat the universe would just periodically oscillate around the center
point; the eventual asymmetric emergence is induced by a
non-trivial change in the underlying potential.)  This raises the
question of what happens to these models when treated quantum
mechanically.  Taking the case of a negative cosmological constant,
corresponding to a constant negative potential, as
a guide suggests that, even though in the flat regions of the
potential there is a classical symmetry expressed by the exact
periodicity in the dynamics, we nevertheless acquire a quantum
mechanical asymmetry due to the evolution in squeezing. A more
complicated question is what happens in the regions where there is
already a classical asymmetry induced by the non-flat potential.

To be specific, let us look at the closed model with $x=0$, while
including a scalar potential $W(\phi)$. In this case, $\phi$ will no
longer serve as a global internal time, but it is still a good
indicator of local internal time in phases where $\phi$ is monotonic
(i.e.\ outside zeros of $p_{\phi}$). In this way, we can still draw
conclusions for the behaviour of quantum variables near a recollapse.
In this case we have the Hamiltonian
\begin{equation}
  H=V\sqrt{P^2+f_0^2 - 8\pi \gamma GW(\phi) f_0^3 V/3} ,
\end{equation}
and a corresponding quantum Hamiltonian
\begin{eqnarray} \label{HQPot}
 H_Q &=& V\sqrt{P^2+f_0^2 - 8\pi \gamma GW(\phi) f_0^3 V/3}+ 
\frac{1}{2}
\frac{V(f_0^2-8\pi\gamma GW(\phi)f_0^3V/3)}{(P^2+f_0^2- 
8\pi\gamma GW(\phi)f_0^3V/3)^{3/2}} G^{PP}\nonumber\\
&&+
\frac{P(P^2+f_0^2- 4\pi\gamma GW(\phi)f_0^3V/3)}{(P^2+f_0^2
- 8\pi\gamma GW(\phi)f_0^3V/3)^{3/2}} G^{VP}\\
&&-\frac{4\pi\gamma G W(\phi)f_0^3(P^2+f_0^2- 
2\pi\gamma GW(\phi)f_0^3V)/3}{(P^2+f_0^2
- 8\pi\gamma GW(\phi)f_0^3V/3)^{3/2}} G^{VV}, \nonumber
\end{eqnarray}
expanded to moments of second order.  In contrast to the previous cases,
this includes not only the quantum moments $G^{PP}$, $G^{VP}$ but also
$G^{VV}$ in correction terms.  The quantum Hamiltonian then determines
equations of motion, which for $G^{VP}$ results in
\begin{eqnarray}
 \frac{\md G^{VP}}{\md\phi} &=& \frac{V(f_0^2-8\pi\gamma
GW(\phi)f_0^3V/3)}{(P^2+f_0^2 -8\pi\gamma G W(\phi)f_0^3V/3)^{3/2}}
G^{PP}\nonumber\\ && +\frac{8\pi\gamma G
W(\phi)f_0^3(P^2+f_0^2- 2\pi\gamma GW(\phi)f_0^3V)/3}{(P^2+f_0^2 -
8\pi\gamma GW(\phi)f_0^3V/3)^{3/2}} G^{VV}\,. \label{GVPeomPot}
\end{eqnarray}
Since $P^2+f_0^2 - 8\pi\gamma GW(\phi)f_0^3V/3$ is required to be
positive and $P$ is small near a recollapse, the sign of this
expression remains unchanged compared to the free model. Thus,
inclusion of a potential does not change our monotonicity
result. Notice that we have not assumed the potential to be small
since the analysis involves only an expansion in moments rather than
in $W(\phi)$.  The rate of change of correlations depends on the value
of the potential, but it has a definite sign: $G^{VP}$ is either
growing or decreasing during a recollapse phase. A varying potential
will affect the rate by which $G^{VP}$ changes, and thus lead to
different absolute changes in squeezing before and after the
recollapse. But correlations will always change, and thus our
qualitative discussion remains unchanged in this case.

\begin{figure}
\begin{center}
\includegraphics[width=14cm]{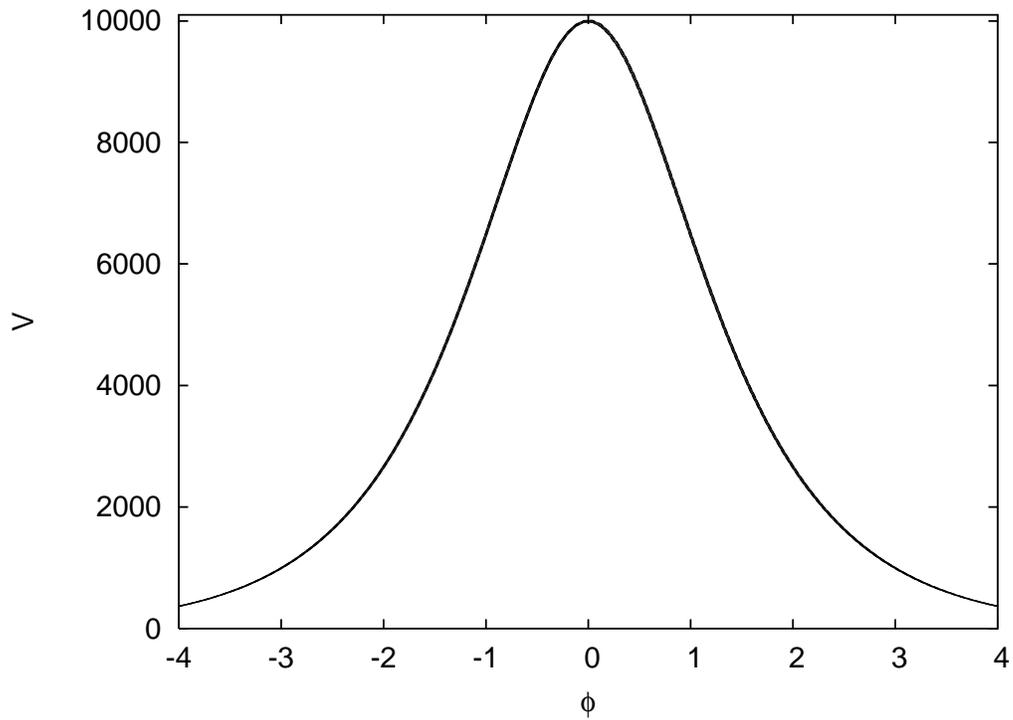}
\caption{\label{EffRecollapse} An example of a recollapsing universe
  which grows to large volume. Plotted are the volume expectation
  value $\langle\hat{V}\rangle(\phi)$ as well as the fluctuations
  around it. While the detailed behaviour of the fluctuations cannot
  be discerned from this total plot, the asymmetry around the
  recollapse is clearly visible in the zoom shown in
  Fig.~\ref{EffRecollapseSmall}.}
\end{center}
\end{figure}

In the cyclic models with many cycles one can only draw conclusions
from the consideration of generic rather than special initial states.
Thus one needs to consider the consequences of generic initially
squeezed states, rather than special unsqueezed initial states.  While
a state may have been uncorrelated at some time, we cannot know how
many cycles ago this may have been, or after how many cycles it may be
so in the future. For statements relevant to a single cycle, which are
the only ones with a chance of being observable, it is not legitimate
to use special initial states which are known to change between
cycles.  In fact as can be seen from Eqs.~(\ref{GVPVLambda}) or
(\ref{GVPVClosed}), there is no strong bound on the change of volume
fluctuations relative to volume from one cycle to the next without a
sharp limit on correlations. Quantum properties of the collapse
  phase can thus differ from those of the expansion phase. As
  (\ref{GVVsol}) shows, the time-asymmetric term has a single factor
  of $V_0$, while the last term is multiplied with $V_0^2$. One can
  thus expect that the asymmetry is not pronounced strongly for a
  universe of large recollapse size $V_0$, but the precise behaviour
  depends also on the moments. Then, the last term containing $V_0^2$
  is suppressed by a factor of $G_0^{PP}$ which must be small near the
  recollapse where $P=0$; see Fig.~\ref{EffRecollapse} for a numerical
  example. Moreover, over several cycles the change in quantum
  properties will add up.

\begin{figure}
\begin{center}
\includegraphics[width=14cm]{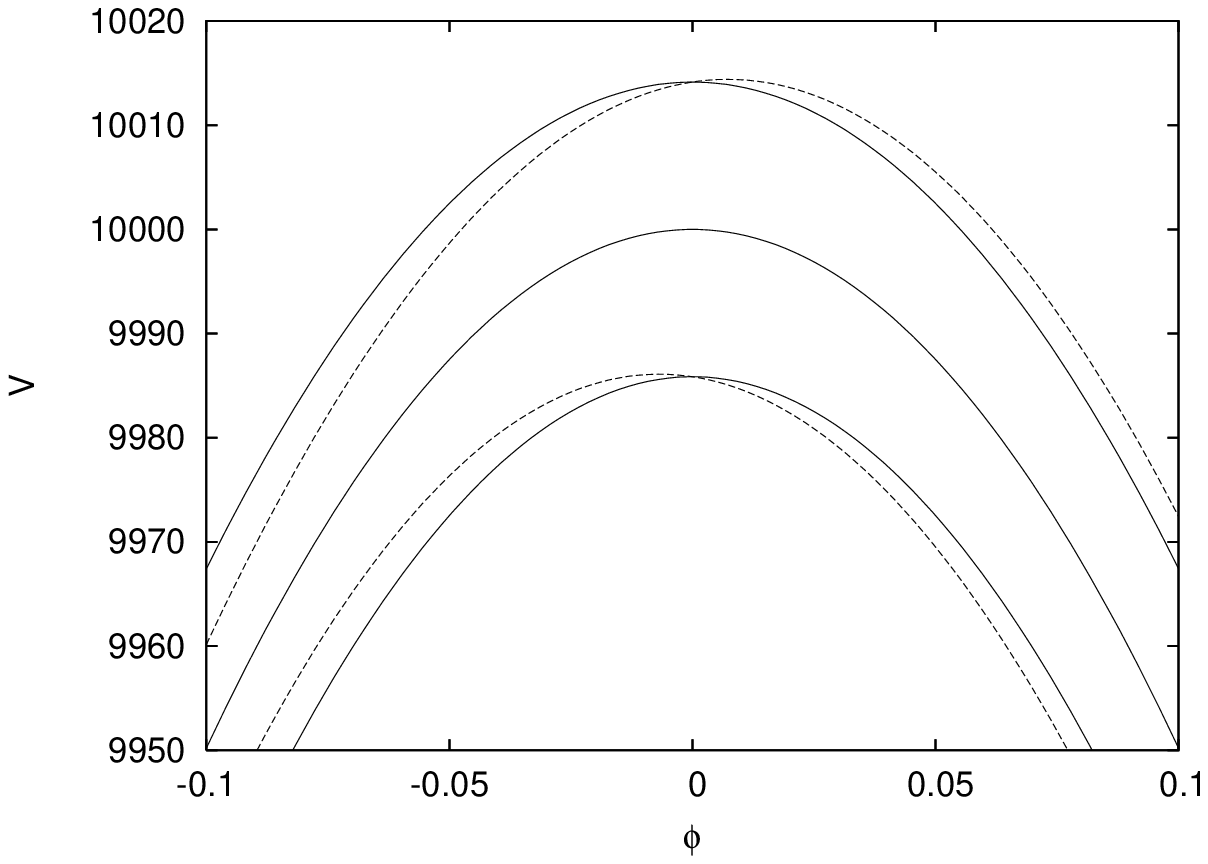}
\caption{\label{EffRecollapseSmall} The recollapse phase of
  Fig.~\ref{EffRecollapse} in more detail. The central line is the
  volume expectation value, and solid lines around it illustrate the
  spread $\Delta V$ of a state which is symmetric around the bounce.
  Dashed lines show how asymmetric the volume fluctuations can be if
  the state is correlated at the recollapse. Initial conditions are
  set at the recollapse, where in units with $\hbar=0.2$ we have
  $V_0=10^4$, $G_0^{VV}=200$, $G_0^{PP}=10^{-4}$ and $G_0^{VP}=0$ for
  the solid lines and $G_0^{VP}=0.1$ for the dashed lines. The latter
  state thus saturates the generalized uncertainty relation (\ref{uncert}).}
\end{center}
\end{figure}

Correlations in a semiclassical state are bounded, and so $G^{VP}$ is
restricted but may certainly vary. And as long as it can easily be
non-zero and affect the behaviour of single cycles, it must be taken
into account in cyclic models with many cycles. Moreover, in addition
to the recollapse phases discussed here, squeezing has a similar
influence on the asymmetry of fluctuations around the bounce
\cite{BounceCohStates}.  This shows that the different cycles of a
universe can indeed be very different from each other, even though
they are connected by deterministic evolution of an underlying state.
The generic behaviour of quantum properties is much more subtle
  than the assumption of unsqueezed states would suggest. Current
knowledge is insufficient to determine what came before, or what will
come after.
\subsection{Entropy}
A central question in cosmology is how to successfully define a notion
of cosmological entropy, and a number of attempts have been made in
this direction. This is in turn hoped to provide a notion of
cosmological arrow of time. The notion of entropy is connected to that
of information associated with the degrees of freedom considered.  In
addition to the usual thermodynamic entropy which is normally
associated with matter/energy degrees of freedom of the constituent
components of the universe \cite{Tolman}, possible notions of
entropy associated with the geometrical \cite{PenroseEntropy} and
gravitational \cite{GravEntro} degrees of freedom of the
universe have also been put forward.

Motivated by the thermodynamical notion of entropy and the associated
second law of thermodynamics a necessary, but not sufficient,
condition that has often been required of general notions of
entropy is that of monotonicity in time. An important step has
therefore often been to look for variables defined in terms of the
underlying dynamics that evolve monotonically.  In addition to notions
of entropy associated with classical degrees of freedom, one would
also expect entropic measures associated with the quantum mechanical
degrees of freedom.  An immediate question that any such measure needs
to answer concerns the nature of its relationship with the
thermodynamic measure of entropy. In particular an important question
in the case of recollapsing/oscillating cosmological models is: do (or
should) the expanding and recollapsing evolutionary phases possess
oscillating or monotonic entropies? Furthermore, how should the
entropy associated with different cycles evolve in oscillating models?

This question has in fact been the subject of a long standing and
intense debate concerning the relation between the so called
thermodynamical and cosmological arrows of time
\cite{Arrow,PenroseEntropy}.  The question is whether the observed
asymmetric (monotonic) thermodynamical time arrow in the current
expanding phase of the universe has a counterpart in cosmology,
particularly in a recollapsing universe. A number of studies have been
made in this connection
\cite{HawkingArrow,EntropyRecollapse,ArrowBoundary}. Given the absence
of a dynamical explanation for the observed asymmetry in the universe,
most such studies assume that the observed thermodynamic arrow of time
must arise from the boundary conditions of the universe
\cite{ArrowBoundary}.

Our results above seem to indicate that the degree of the squeeze of
the quantum gravity state may provide a notion of entropy purely
associated with quantum degrees of freedom To the best of our
knowledge, this is a new possibility not considered before. (Relating
entropy to the squeezing of a matter state, however, has been
considered in the context of particle production; see e.g.\
\cite{GasperiniEntropy,LargeSqueezing,EntropyOpen,TimeSymmGeomPhase,CosmoEntroProd,RecollEntroProd}.)
As can be seen from (\ref{GVPeom}) and (\ref{GVPeomPot}), the
squeezing of a state is strictly monotonic in time during expansion,
recollapse and contraction of a cycle in the models considered.  This
demonstrates that even in isotropic models, which include the
microscopic dynamics only in a highly averaged form, quantum aspects
prevent one from viewing a collapsing universe simply as a
time-reverse of its expansion. The quantum theory's arrow of time
cannot reverse at the recollapse.

Unfortunately, it is difficult to follow its evolution through a
bounce because this phase can only be described in a different set of
basic variables ($J=V\exp(iP)$ for $P$) which make the equations
solvable. For classical variables, these are easily translated into
each other. But the transformation is non-linear, such that moments
transform in a highly complicated way. In any case, the change in
squeezing is nevertheless generic because it is unlikely that the
bounce will restore fluctuations to precisely the value of the
preceding cycle.  Moreover, one can roughly estimate the squeezing as
it evolves through the bounce. In the bounce phase, only operators
such as $\hat{J}:=\hat{V}\exp(i\hat{P})$ exist and give rise to a
solvable evolution. Moments between $V$ and $J$ can thus be computed
exactly \cite{BounceCohStates}, but it is difficult to transform
between the $V$-$J$ and $V$-$P$ moments. However, the bounce happens near
$P\approx\pi/2$, and with $\delta P:=P-\pi/2$ we have, up to
reordering, ${\rm
  Re}J=\langle\hat{V}\cos(\hat{P})\rangle=
\langle\hat{V}\cos(\pi/2+\delta \hat{P})\rangle=
-\langle\hat{V}\sin\delta \hat{P}\rangle\sim -\langle\hat{V}\delta
\hat{P}\rangle= -\langle\hat{V}(\hat{P}-\pi/2)\rangle$. Thus, ${\rm
  Re}J+VP-\frac{\pi}{2}V\sim -G^{VP}$ provides an estimate for the
$V$-$P$ squeezing as it evolves through the bounce in terms of
expectation values. Since expectation values are symmetric around the
bounce in the absence of a potential, not much additional squeezing is
generated around the bounces. 

Most of the squeezing is thus generated in the recollapse phases,
  which resembles recent results for cyclic models with bounces based
  on the Hagedorn phase of string theory \cite{CyclicHagedorn}. In the
  present context with a quantum measure for entropy in the form of
  squeezing, this may seem counterintuitive given that the
recollapse is a much more classical phase than the bounce.  However,
the production of correlations is not so much a matter of quantum
versus semiclassical behaviour but rather of the dynamics in a given
regime. A state may remain semiclassical to an excellent degree, and
yet receive a significant amount of squeezing. Whether or not this
happens depends on the equation of motion for $G^{VP}$, or the
underlying Hamiltonian.  The analysis presented in this article
unambiguously shows the production of correlations in a recollapse
phase even though it is semiclassical. Although our qualitative
estimates for the bounce phases are difficult to make precise, the
monotonic behaviour of correlations at small curvature appears to be
an interesting and reliable property.

The precise amount of squeezing depends on initial conditions. If all
moments could initially be zero, they would remain so and no squeezing
would develop. However, this initial condition is impossible because
the moments are subject to the uncertainty relation (\ref{uncert}).
Thus, unless the volume uncertainty diverges, $G^{PP}$ cannot be zero
in (\ref{GVPeom}) and an initially unsqueezed state inevitably
develops squeezing over time which can grow large over many cycles. It
is thus quantum uncertainty, together with the specific dynamics of
the system, which prevents the existence of perfectly symmetric
states.

There is a sense in which small squeezing presents a special state
with a distinguished discrete symmetry. Under time reversal, we map
$\phi\mapsto -\phi$, $P\mapsto -P$ and $G^{VP}\mapsto -G^{VP}$ while
the other variables remain unchanged. Thus, a time reversible solution
would have vanishing squeezing which one may view as a special state
analogous to low entropy. As (\ref{GVVsol}) shows, this is
  obtained for vanishing correlations at the recollapse. However,
since $G^{VP}$ would generically be non-zero at a recollapse,
  especially in a cyclic model, there is no solution which is exactly
time reversible.  Again, it is the uncertainty relation as an
additional condition, which eliminates those initial values which
would correspond to time reversal solutions.

\section{Conclusions}
We have studied the evolution of recollapsing models within an
isotropic and homogeneous quantum cosmological framework in presence
of a scalar field. To allow a recollapse we consider, in turn, a
negative cosmological constant as well as a positive curvature model.
We derive the resulting quantum evolution equations to second order in
moments of a state and study their effects on the recollapsing
dynamics of the universe, i.e. the expanding, turn around and
contracting phases. These effective equations allow us to observe that
state properties generically change during the recollapse, making
quantum fluctuations in the expansion and contraction phases
different. At large volumes as they are realized at a recollapse, the
change is not as noticeable as it can be for states travelling through
a bounce \cite{BeforeBB,Harmonic}, but it is significant especially in
a cyclic model with several recollapse phases. As in the case of the
bounce, the asymmetry of fluctuations is controlled by quantum
correlations which have often been ignored in previous studies.

The specific equations analyzed here thus allow us to identify
correlations as a quantum measure for the change of fluctuations.
More precisely, we find that the squeezing of an initial state is
strictly monotonic in time throughout these three phases for the
models considered. Importantly, we have shown this finding to be
robust under the inclusion of a matter potential. Combining these
results with the corresponding ones concerning a bounce in loop
quantum cosmology we have shown that squeezing of an initial state
evolves monotonically throughout a whole cycle.  The absence of
perfectly symmetric states is a combined consequence of the specific
dynamics of the quantum system together with the presence of quantum
uncertainty.

Such monotonicity is of potential importance in two regards.  Firstly,
it sheds new light on a long standing intensive debate concerning the
(a)symmetry between the expanding and contracting phases in a
recollapsing universe. As shown here, the contracting phase cannot be
a time reverse of the expanding phase.  Secondly, it motivates the
adoption of the degree of squeezing as an alternative measure of
(quantum) entropy.

Qualitatively, we also consider the evolution of the squeezing of an
initial state in emergent nonsingular oscillating universes in which
the universe is assumed to have undergone a large (possibly infinite)
number of past cycles. We argue that the consideration of the amount
of squeezing in the universe can in principle provide some constraints
on the viability of such emergent models. In any case, given that a
generic cycle does have non-vanishing correlations, squeezings of the
quantum gravity state must be taken into account in order to draw
reliable conclusions about cyclic models.
\section*{Acknowledgements}
This work was supported in part by NSF grant PHY0653127.


\begin{thebibliography}{10}

\bibitem{LivRev}
M.\ Bojowald,
\newblock Loop Quantum Cosmology,
\newblock {\em Living Rev.\ Relativity} 8 (2005) 11, [gr-qc/0601085],
\newblock {\tt http://relativity.livingreviews.org/Articles/lrr-2005-11/}

\bibitem{BSCG}
M.\ Bojowald,
\newblock Singularities and Quantum Gravity,
\newblock {\em AIP Conf.\ Proc.} 910 (2007) 294--333,
  [gr-qc/0702144],
\newblock In: Proceedings of the XIIth Brazilian School on Cosmology and
  Gravitation

\bibitem{PreBigBang}
M.\ Gasperini and G.\ Veneziano,
\newblock The pre-big bang scenario in string cosmology,
\newblock {\em Phys.\ Rept.} 373 (2003) 1--212

\bibitem{Ekpyrotic}
J.\ Khoury, B.~A.\ Ovrut, P.~J.\ Steinhardt, and N.\ Turok,
\newblock The Ekpyrotic Universe: Colliding Branes and the Origin of the Hot
  Big Bang,
\newblock {\em Phys.\ Rev.\ D} 64 (2001) 123522, [hep-th/0103239]

\bibitem{CyclicEkpy}
P.~J.\ Steinhardt and N.\ Turok,
\newblock Cosmic Evolution in a Cyclic Universe,
\newblock {\em Phys.\ Rev.\ D} 65 (2002) 126003, [hep-th/0111098]

\bibitem{CyclicDesign}
J.\ Khoury, P.~J.\ Steinhardt, and N.\ Turok,
\newblock Designing cyclic universe models,
\newblock {\em Phys.\ Rev.\ D} 92 (2004) 031302

\bibitem{QuantumBigBang}
A.\ Ashtekar, T.\ Pawlowski, and P.\ Singh,
\newblock Quantum Nature of the Big Bang,
\newblock {\em Phys.\ Rev.\ Lett.} 96 (2006) 141301, [gr-qc/0602086]

\bibitem{APS}
A.\ Ashtekar, T.\ Pawlowski, and P.\ Singh,
\newblock Quantum Nature of the Big Bang: An Analytical and Numerical
  Investigation,
\newblock {\em Phys.\ Rev.\ D} 73 (2006) 124038, [gr-qc/0604013]

\bibitem{APSII}
A.\ Ashtekar, T.\ Pawlowski, and P.\ Singh,
\newblock Quantum Nature of the Big Bang: Improved dynamics,
\newblock {\em Phys.\ Rev.\ D} 74 (2006) 084003, [gr-qc/0607039]

\bibitem{APSCurved}
A.\ Ashtekar, T.\ Pawlowski, P.\ Singh, and K.\ Vandersloot,
\newblock Loop quantum cosmology of $k=1$ FRW models,
\newblock {\em Phys.\ Rev.\ D} 75 (2007) 024035, [gr-qc/0612104]

\bibitem{BouncePert}
M.\ Bojowald,
\newblock Large scale effective theory for cosmological bounces,
\newblock {\em Phys.\ Rev.\ D} 75 (2007) 081301(R), [gr-qc/0608100]

\bibitem{BounceCohStates}
M.\ Bojowald,
\newblock Dynamical coherent states and physical solutions of quantum
  cosmological bounces,
\newblock {\em Phys.\ Rev.\ D} 75 (2007) 123512, [gr-qc/0703144]

\bibitem{BouncePot}
M.\ Bojowald, H.\ Hern\'andez, and A.\ Skirzewski,
\newblock Effective equations for isotropic quantum cosmology including matter,
\newblock {\em Phys.\ Rev.\ D} 76 (2007) 063511, [arXiv:0706.1057]

\bibitem{QuantumBounce}
M.\ Bojowald,
\newblock Quantum nature of cosmological bounces, arXiv:0801.4001

\bibitem{VaasLQCII}
M.\ Bojowald and R.\ Tavakol,
\newblock Loop Quantum Cosmology II: Effective theories and oscillating
  universes, In R.\ Vaas, editor, {\em Beyond the Big Bang},
\newblock Springer, Berlin, 2008, [arXiv:0802.4274]

\bibitem{TolmanEntropy}
R.~C.\ Tolman,
\newblock On the problem of entropy of the universe as a whole,
\newblock {\em Phys.\ Rev.} 37 (1931) 1639--1660

\bibitem{Tolman}
R.~C.\ Tolman,
\newblock {\em Relativity, Thermodynamics and Cosmology},
\newblock Clarendon Press, Oxford, 1934

\bibitem{BeforeBB}
M.\ Bojowald,
\newblock What happened before the big bang?,
\newblock {\em Nature Physics} 3 (2007) 523--525

\bibitem{Harmonic}
M.\ Bojowald,
\newblock Harmonic cosmology: How much can we know about a universe before the
  big bang?,
\newblock {\em Proc.\ Roy.\ Soc.\ A} (2008) to appear, [arXiv:0710.4919]

\bibitem{BounceSqueezed}
M.\ Bojowald,
\newblock in preparation

\bibitem{InhomLattice}
M.\ Bojowald,
\newblock Loop quantum cosmology and inhomogeneities,
\newblock {\em Gen.\ Rel.\ Grav.} 38 (2006) 1771--1795, [gr-qc/0609034]

\bibitem{SchwarzN}
M.\ Bojowald, D.\ Cartin, and G.\ Khanna,
\newblock Lattice refining loop quantum cosmology, anisotropic models and
  stability,
\newblock {\em Phys.\ Rev.\ D} 76 (2007) 064018, [arXiv:0704.1137]

\bibitem{QSDI}
T.\ Thiemann,
\newblock Quantum Spin Dynamics {(QSD)},
\newblock {\em Class.\ Quantum Grav.} 15 (1998) 839--873, [gr-qc/9606089]

\bibitem{EffAc}
M.\ Bojowald and A.\ Skirzewski,
\newblock Effective Equations of Motion for Quantum Systems,
\newblock {\em Rev.\ Math.\ Phys.} 18 (2006) 713--745, [math-ph/0511043]

\bibitem{Karpacz}
M.\ Bojowald and A.\ Skirzewski,
\newblock Quantum Gravity and Higher Curvature Actions,
\newblock {\em Int.\ J.\ Geom.\ Meth.\ Mod.\ Phys.} 4 (2007) 25--52,
  [hep-th/0606232],
\newblock In: Proceedings of ``Current Mathematical Topics in Gravitation and
  Cosmology'' (42nd Karpacz Winter School of Theoretical Physics), Ed.\
  Borowiec, A.\ and Francaviglia, M.

\bibitem{BounceSols}
J.\ Mielczarek, T.\ Stachowiak, and M.\ Szyd\l{}owski,
\newblock Exact solutions for big bounce in loop quantum cosmology,
  arXiv:0801.0502

\bibitem{Sing}
M.\ Bojowald,
\newblock Absence of a Singularity in Loop Quantum Cosmology,
\newblock {\em Phys.\ Rev.\ Lett.} 86 (2001) 5227--5230, [gr-qc/0102069]

\bibitem{HomCosmo}
M.\ Bojowald,
\newblock Homogeneous loop quantum cosmology,
\newblock {\em Class.\ Quantum Grav.} 20 (2003) 2595--2615, [gr-qc/0303073]

\bibitem{Spin}
M.\ Bojowald, G.\ Date, and K.\ Vandersloot,
\newblock Homogeneous loop quantum cosmology: The role of the spin connection,
\newblock {\em Class.\ Quantum Grav.} 21 (2004) 1253--1278, [gr-qc/0311004]

\bibitem{SphSymmSing}
M.\ Bojowald,
\newblock Non-singular black holes and degrees of freedom in quantum gravity,
\newblock {\em Phys.\ Rev.\ Lett.} 95 (2005) 061301, [gr-qc/0506128]

\bibitem{Rov}
C.\ Rovelli,
\newblock {\em Quantum Gravity},
\newblock Cambridge University Press, Cambridge, UK, 2004

\bibitem{ALRev}
A.\ Ashtekar and J.\ Lewandowski,
\newblock Background independent quantum gravity: A status report,
\newblock {\em Class.\ Quantum Grav.} 21 (2004) R53--R152, [gr-qc/0404018]

\bibitem{ThomasRev}
T.\ Thiemann,
\newblock {\em Introduction to Modern Canonical Quantum General Relativity},
\newblock Cambridge University Press, Cambridge, UK, 2007

\bibitem{Emergent}
G.~F.~R.\ Ellis and R.\ Maartens,
\newblock The Emergent Universe: inflationary cosmology with no singularity,
\newblock {\em Class.\ Quant.\ Grav.} 21 (2004) 223--232, [gr-qc/0211082]

\bibitem{Emergent2}
G.~F.~R.\ Ellis, J.\ Murugan, and C.~G.\ Tsagas,
\newblock The Emergent Universe: An Explicit Construction,
\newblock {\em Class.\ Quant.\ Grav.} 21 (2004) 233--250, [gr-qc/0307112]

\bibitem{EmergentLoop}
D.~J.\ Mulryne, R.\ Tavakol, J.~E.\ Lidsey, and G.~F.~R.\ Ellis,
\newblock An emergent universe from a loop,
\newblock {\em Phys.\ Rev.\ D} 71 (2005) 123512, [astro-ph/0502589]

\bibitem{EmergentNat}
M.\ Bojowald,
\newblock Original Questions,
\newblock {\em Nature} 436 (2005) 920--921

\bibitem{EinsteinStaticHol}
L.\ Parisi, M.\ Bruni, R.\ Maartens, and K.\ Vandersloot,
\newblock The Einstein static universe in Loop Quantum Cosmology,
\newblock {\em Class.\ Quantum Grav.} 24 (2007) 6243--6253, [arXiv:0706.4431]

\bibitem{PenroseEntropy}
R.\ Penrose,
\newblock In {\em General Relativity: An Einstein Centenary Survey},
\newblock Cambridge University Press, 1979

\bibitem{GravEntro}
F.~C.\ Mena and R.\ Tavakol,
\newblock Evolution of the density contrast in inhomogeneous dust models,
\newblock {\em Class.\ Quantum Grav.} 16 (1999) 435--452

\bibitem{Arrow}
T.\ Gold,
\newblock The arrow of time,
\newblock {\em Am.\ J.\ Phys.} 30 (1962) 403--410

\bibitem{HawkingArrow}
S.~W.\ Hawking,
\newblock Arrow of time in cosmology,
\newblock {\em Phys.\ Rev.\ D} 32 (1985) 2489--2495

\bibitem{EntropyRecollapse}
D.~N.\ Page,
\newblock Will entropy decrease if the Universe recollapses?,
\newblock {\em Phys.\ Rev.\ D} 32 (1985) 2496--2499

\bibitem{ArrowBoundary}
S.~W.\ Hawking, R.\ Laflamme, and G.~W.\ Lyons,
\newblock Origin of time asymmetry,
\newblock {\em Phys.\ Rev.\ D} 47 (1993) 5342--5356

\bibitem{GasperiniEntropy}
M.\ Gasperini and M.\ Giovannini,
\newblock Quantum squeezing and cosmological entropy production,
\newblock {\em Class.\ Quantum Grav.} 10 (1993) L133--L136

\bibitem{LargeSqueezing}
M.\ Kruczenski, L.~E.\ Oxman, and M.\ Zaldarriaga,
\newblock Large squeezing behaviour of cosmological entropy generation,
\newblock {\em Class.\ Quantum Grav.} 11 (1994) 2317--2329

\bibitem{EntropyOpen}
D.\ Koks, A.\ Matacz, and B.~L.\ Hu,
\newblock Entropy and uncertainty of squeezed quantum open systems,
\newblock {\em Phys.\ Rev.\ D} 55 (1997) 5917--5935,
\newblock Erratum: \cite{EntropyOpenErr}

\bibitem{TimeSymmGeomPhase}
S.~P.\ Kim and S.-W.\ Kim,
\newblock Will geometric phases break the symmetry of time in quantum
  cosmology?,
\newblock {\em Phys.\ Rev.\ D} 49 (1994) R1679--R1683

\bibitem{CosmoEntroProd}
S.~P.\ Kim and S.-W.\ Kim,
\newblock Quantum cosmological entropy production and the asymmetry of
  thermodynamic time,
\newblock {\em Phys.\ Rev.\ D} 51 (1995) 4254--4258

\bibitem{RecollEntroProd}
S.~P.\ Kim and S.-W.\ Kim,
\newblock Entropy production and thermodynamic arrow of time in a recollapsing
  universe,
\newblock {\em Nuovo Cim.\ B} 115 (2000) 1039--1048

\bibitem{CyclicHagedorn}
T.\ Biswas,
\newblock The Hagedorn Soup and an Emergent Cyclic Universe, arXiv:0802.0176

\bibitem{EntropyOpenErr}
D.\ Koks, A.\ Matacz, and B.~L.\ Hu,
\newblock {\em Phys.\ Rev.\ D} 56 (1997) 5281

\end{thebibliography}

\end{document}